# Validation of the IS Impact Model for Measuring the Impact of e-Learning Systems in KSA Universities: Student Perspective

Salem Alkhalaf, Steve Drew, Anne Nguyen

School of ICT, Griffith University,
Gold Coast, Australia

*Abstract*—**The IS-Impact Measurement Model, developed by Gable, Sedera and Chan in 2008, represents the to-date and expected stream of net profits from a given information system (IS), as perceived by all major user classes. Although this model has been stringently validated in previous studies, its generalizability and verified effectiveness are enhanced through this new application in e-learning. This paper focuses on the re-validation of the findings of the IS-Impact Model in two universities in the Kingdom of Saudi Arabia (KSA). Among the users of 2 universities e-learning systems, 528 students were recruited. A formative validation measurement with SmartPLS, a graphical structural equation modeling tool was used to analyse the collected data. On the basis of the SmartPLS results, as well as with the aid of data-supported IS impact measurements and dimensions, we confirmed the validity of the IS-Impact Model for assessing the effect of e-learning systems in KSA universities. The newly constructed model is more understandable, its use was proved as robust and applicable to various circumstances.**

## I. INTRODUCTION

According to the Communications and Information Technology Commission (CITC), the Kingdom of Saudi Arabia (KSA) is one of the fastest growing countries in the world in terms of e-learning [1]. CITC data shows an explosive growth in the number of Internet users generally, from a mere 200,000 in 2000 to 4.8 million in 2006. The number of students enrolled in Saudi institutions of higher education has also increased significantly in 2011 to 11.4 million [1]. As a result, many of these institutions have turned to e-learning systems as a means to help broaden and enhance access to their courses and subjects [2].

Reflecting this trend, a growing number of research studies have been conducted on e-learning in the Kingdom of Saudi Arabia [3-5]. Movement and development in e-learning seem to be fast and strong, especially in the universities in KSA [3].

Many of these studies have focused on identifying the key factors that differentiate online education from face-to-face learning, analysing the in-principle advantages and disadvantages of online courses, or developing strategies to achieve a suitable online learning environment [6]. To date, however, little attention has been paid to the issue of assessing the e-learning environments that have been set up in the country. Indeed, it would appear that relatively little research has been done regarding the evaluation of e-Learning systems in general [7, 8].

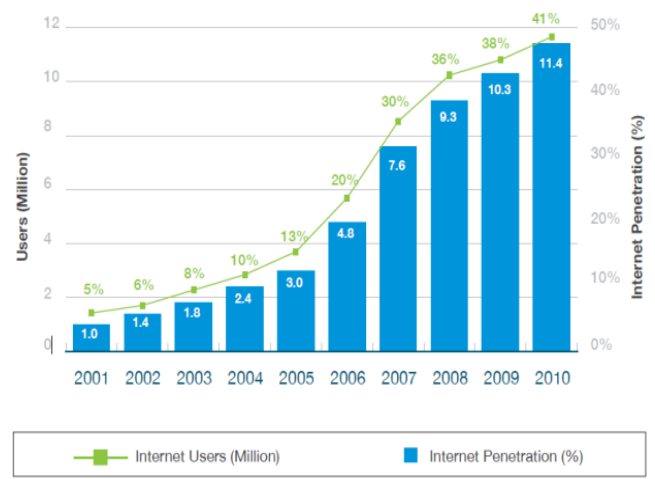

Figure 1.   Internet Market Evolution for KSA (2001-2010) (adopted from [1])

## II. BACKGROUND

Interest in e-learning has grown rapidly during the past decade or so in KSA, for a number of reasons [9]. First, the demand for higher education has far outstripped supply, such that institutions are faced with overcrowding and insufficiency of facilities and human resources for the delivery of traditional-style education to all of the nation's qualified applicants for admission. E-learning has been suggested as a means to overcome these limitations.

Second, KSA is a large country in terms of geographical area, with a significant number of communities being isolated from major population centres. E-learning offers the potential to deliver educational services to remote locations, thereby reducing disparities across the various regions and areas [5, 10].

Third, in KSA's higher education, men and women receive their instruction in separate classes not mixed "ikhtilat" classes, for cultural and religious reasons [11]. So, Male instructors can only teach female students through eLearning tools or distance learning tools [10]. This puts further strains on the limited facilities and human resources available. It has been observed, accordingly, that women are often among the strongest supporters of e-learning, which potentially facilitates their access to higher education [12-14].

In 2008 the KSA Ministry of Higher Education established a National Centre of E-learning & Distance Learning to promote and facilitate the spread of e-learning systems in Saudi universities [15]. It has was estimated that in 2008, annual turnover of the e-learning industry in KSA had already reached US$ 125 million, with further expected growth of about 33 per cent per year over the following 5 years [16]. It is timely, therefore, to investigate the issue of assessing the success or impact of the e-learning systems that have been set up to date[5].

### III. CONCEPTUAL FRAMEWORK AND MEASUREMENT DIMENSIONS

The IS-Success/Impact Measurement framework was selected because it comprehensively takes into account four dimensions of success/impact in the context of IT systems [17]. The IS-Success Model [18] is one of the most cited models in IS research [19]. More recently it has been supplemented by the IS-Impact Measurement Model [20]. As illustrated in Figure II, within the IS-Success/Impact framework, the success and impact of an IS system can be measured in terms of

- the quality of the information produced (information quality),

- the performance of the system from a technical perspective (system quality),

- the impact on individual users (individual impact), and

- the impact on the relevant organisation (organisational impact).

For e-learning systems, the third and fourth dimensions are the most important ones, as they represent the end-goals of the system [5].

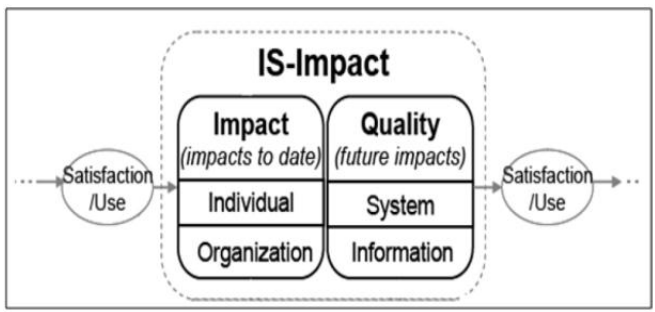

Figure 2. IS-Impact Model (adapted from [20])

Gable, Sedera, & Chan [20] stated that on behalf of an organisation, Individual Impact measures the extent to which an IS influences the capabilities and effectiveness of key stakeholders or users [21]. Organisational Impact is a measure of the degree to which the IS advances the enhancement of organisational outcomes and abilities. Information Quality is a measure of the quality of (the IS) output (e.g., the quality of the information that the system produces in reports and onscreen) [20]. System Quality refers to a measure of (the IS) performance from a technical and design perspective [20].

The IS-Impact Model is considered the theoretical foundation for this study. We designed the model in such a way that it is robust and generalisable, and produces simple results that are highly comparable across stakeholders, time, and various types of systems [21]. We employ a continuous measurement scheme, thereby creating a general instrument that is easy to understand and respond to, by all relevant stakeholder groups. This measurement scheme enables the comparison of stakeholder perspectives.

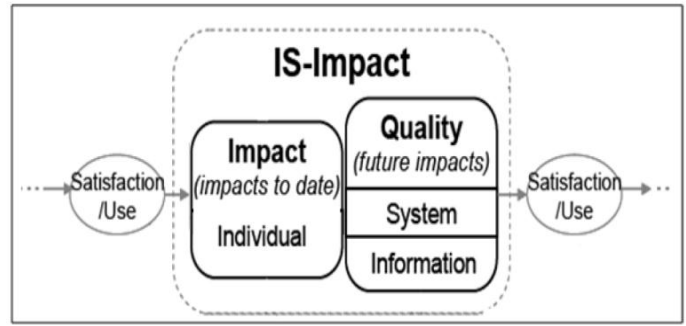

Figure 3. New measurement model for students using e-learning systems

The present study identifies relevant new measures that were not determined by previous researchers. In order to concentrate on the impact that e-learning systems have on students, as the key stakeholders, for the purposes of this paper we exclude the measure of "Organizational Impact". Thus, we use only 26 measures (the a priori model developed in [20]) to ensure model completeness.

### IV. RESEARCH METHODOLOGY

To ensure the suitability of the model for KSA universities, we modified the original survey instrument. It was translated into Arabic for students who are less conversant in English. Following the suggestions of Brislin and McGorry [21, 22], the instrument was translated using both back translation and decentering approaches. Sentence structure and word choice were slightly modified, ensuring that no changes in meaning occurred [21].

The Arabic version was then examined for face validity. Given that the Organisational Impact dimension is more strongly related to staff and faculty member issues than to student attributes, we deducted the items that make up this dimension. For staff members, other models such as fourth dimensions were introduced.

The questionnaire comprises two main sections. The first is intended to collect demographic information on the respondents, while the second features the 26 measures of the IS-Impact Model and many dependent variables used to test construct validity. A five-point Likert scale (with "strongly agree" and "strongly disagree" as the end points) was used to rank the responses of the participants regarding the model items [5]. Figure IV shows the IS-Impact Model and its 26 measures.

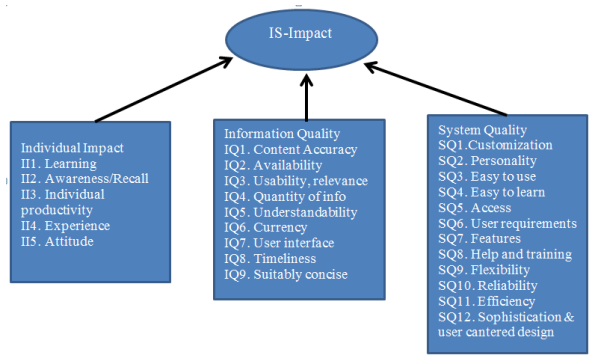

Figure 4.   IS-Impact Model with 26 measures

The participants were required to answer all the questions, which include those related to the descriptive items, IS-Impact measures, and dependent variables. This instruction is written at the beginning of the questionnaire and repeated in the introductory page.

Qassim University and King AbdulAziz University in KSA were chosen as the study sites for data collection. A mixture of cluster, convenience, and snowball sampling methods were used to select the respondents. The respondents belong to e-learning classes or use the e-learning systems of the universities. A hard copy of the questionnaire was distributed to the sample. Out of the 800 questionnaires distributed, 560 were returned; 32 were rejected because of incomplete answers. The final sample comprised 528 students (328 males, 200 females).

## V.   RESULTS

A clear definition and understanding of formative construct and its difference with reflective construct were obtained; in specifying formative constructs, few studies provided exemplary interpretations of formative measurement results, but we were able to find credible ones from [23] and [24]. An illustrative example of formative construct validation was found in [25], [26], and [27]. The authors demonstrated the method for estimating and assessing constructs with partial least squares (PLS) software. One of the principal advantages of PLS is that it supports both formative and reflective measures—a highly beneficial feature because employing covariance-based SEM techniques, such as LISREL or EQS [28, 29], is difficult to accomplish. Assessing the validity of a formative measurement model involves four steps, shown in the Table I below.

Validity test for the formative measurement model (adopted from [21])

| Test of | Description |
|---|---|
| Multicollinearity | The presence of multicollinearity of the items is identified by conducting a test.<br>Higher collinearity among items shows conceptual redundancy. |
| External validity | External validity is used to evaluate the validity by examining how well the formative items capture the construct by relating these measures with a reflective variable of the same |

| | construct. |
|---|---|
| Nomological validity (Nomological net) | Nomological validity is used to evaluate validity by connecting the items to other constructs that have important and strong relationships known through previous research; that is, according to previous, research linking the formative measurement model with the antecedents and/or consequence constructs to which a structural path exists. |
| Significance of weights | Formative measurement models are observed by significant weights. |

### A.   Multicollinearity

Collinearity diagnostics is a regression test on the items in a formative construct (the independent variables) with a dependent variable. The results show the presence of collinearity. The presence of multicollinearity was also confirmed by the tolerance and variation inflation factor (VIF) value shown in the "Coefficients" table. The results demonstrate that all the 26 measures were below the general VIF cutoff point of 10 [23], [30], [31], with 1.9 as the largest VIF observed.

Diamantopolous & Winklhofer [23] stated that one approach to testing item quality is observing the correlation of items with a different variable that may be external to an index. The items that show a strong correlation with the variable should be retained. Further, Diamantopolous & Winklhofer [23] suggested that the relationships among the items with the proposed dependent variable (the global item) at each dimension can be examined by using four global items that "summarise the essence of the construct that the index purports to measure".

### B.   Assessing the Validity of the IS-Impact Model through Structural Relationships

The IS-Impact Model was tested by determining the structural relationship between unobserved variables (inner model), also known as nomological (net) validity [21]. The tests were carried out using SmartPLS, a software application for (graphical) path modelling with latent variables, in which the PLS approach is used to analyse latent variables, or both latent variables and observed or manifest variables (outer model) [21, 32]. Figure V shows the results of the structural relationship evaluation.

Two reflective measures given in the "IS-Impact summary" were used to evaluate the structural relationship of the measurement model (outer model). The analysis shows adjusted R-squares of 0.622 and 0.941, indicating that 94.1% of the variance in the IS-Impact Model is explained by Individual Impact, Information Quality, and System Quality. Chin [33] and Henseler, Ringle, & Sinkovics [27] suggested R-square values of 0.67 (substantial), 0.33 (moderate), and 0.19 (weak).

The IS-Impact Model is almost substantial, as shown by the path analysis. At $p < 0.05$ or better (estimated by a bootstrapping procedure with 528 bootstrapping samples), all the structural paths were significant and the IS impact was considerably influenced by System Quality and Information Quality.

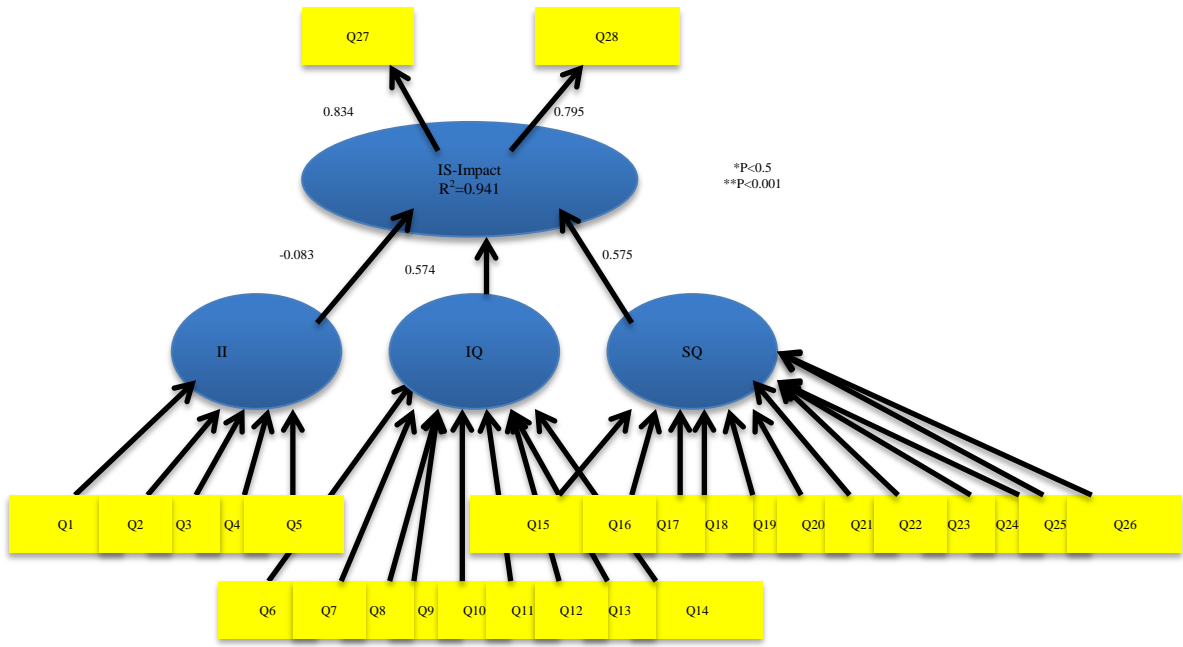

Figure 5. Structural model

## C. Explanatory Power of the Model

After the PLS test, the R-square was varied to determine the effect of Individual Impact, Information Quality, and System Quality on the overarching IS-Impact construct. This approach involves the repeated estimation of PLS and evaluation of effect size, in which each of the PLS is run with one dimension omitted. The effect size was derived using the following equation:

$$\text{Effect size, } f^2 = \frac{R_{included}^2 - R_{excluded}^2}{1 - R_{included}^2} \qquad (1)$$

Effect size (adopted from [21])

Table II shows that Individual Impact, Information Quality, and System Quality exhibited an average effect of greater than 0.15 on IS impact. Cohen [34] denoted a value of 0.02 as small, 0.15 as medium, and 0.35 as large. The effect of System Quality on IS impact was larger at 0.35. Demonstrating the addition of the three dimensions as a complete measurement model is the main objective of this analysis. The results show that mixing all the dimensions in a model enabled a high incremental change in the R-square variations.

TABLE I. EFFECT SIZE

| R-square included | | 0.941 | | |
|---|---|---|---|---|
| Run | Removed | R-square Excluded | Effect Size | Interpretation |
| 1 | Individual impact (II) | 0.937 | 0.068 | Small effect |
| 2 | Information quality (IQ) | 0.919 | 0.322 | Large effect |
| 3 | System quality (SQ) | 0.922 | 0.373 | Medium effect |

## D. Variance Inflation Factor

The VIF factor scores for all the three dimension were low (all VIF < 2), strongly indicating that most of the predictor variables in the model were confounded [35]. Because the values (VIF) less than 10 are the critical values for this factor, no overlap was observed between these factors; they are independent of determinants with a significant overlap. This result is unimportant and can be considered similar to the unexplained change in the follower, but it also indicates no multicollinearity problem [36].

TABLE II. VIF

| Category | VIF |
|---|---|
| Individual impact (II) | 1.9 |
| Information quality (IQ) | 1.5 |
| System quality (SQ) | 1.4 |

## VI. DISCUSSION AND CONCLUSION

The IS-Impact Model can be used to determine the effect of e-learning information systems used to date, as well as to forecast its future effects on KSA universities. IS impact and all the pathways between the construct and higher constructs were significant, suggesting that. The hypotheses below are supported by the Model; therefore, the students were definitely satisfied because of the considerable impact of the IS.

TABLE III. HYPOTHESES

| No | Hypotheses | Results |
|---|---|---|
| H1 | Individual Impact is positively affected by the use of collaborative eLearning | Supported. |
| H2 | System Quality is positively associated with the use of collaborative eLearning system. | Supported. |
| H3 | Information Quality is positively affected by employing a collaborative eLearning system. | Supported. |

We conducted many path estimate tests to exclude the measures with low path coefficients and observe the incremental change in R-square values. When these measures were excluded, we noticed a decrease in R-square values at the end of the tests, indicating that the explanatory power of the model diminishes. Therefore, although some measures may not be strong predictors of the construct, they are still considered to have important correlations in this study. The IS-Impact construct was considerably influenced by all the 26 measures, and none should be excluded.

We addressed the generalisability of the IS-Impact Model and expanded its application to new settings; that is, we used varied languages, cultures, and types of systems. In the context of KSA universities, the Model proves to be a valid tool. Adding or excluding a measure from the original set resulted in a more understandable model for assessing the effect of IS on KSA universities. Both the Arabic and English instruments were also suitable for determining the validity of the Model. Therefore, the IS-Impact Model is rich and can be applied to various circumstances (custom package, KSA universities, and English or Arabic language).

The limitations of this study can serve as opportunities for future research. Although contexts differ from one another, the differences can be minimised by using the Model. The e-learning system was used as a unit of analysis, similar to the original work on the IS-Impact Model. The generalisability of the Model was tested and the validity of the IS evaluation system was expanded to the KSA context. Future studies can widen this scope by testing validity under different contexts. For instance, the assessment of the effect of IS on organisations can be carried out by developing better standardised measurement systems.

Finally, future work would be to developing the IS-Impact Model for assessing the effect of e-learning systems in other educational environments of eLearning. Also focusing on supporting processes of eLearning, the research will be conducted taking into account the general requirements of eLearning system.

AUTHORS' PROFILES

**Salem Alkhalaf** graduated Bachelor of Education degree in Computer Since from the Department of Computer, Teachers College (Riyadh) in 2003, Also he graduated with Honors degree. And he graduated the Master of ICT from Griffith University (Brisbane, Australia) in 2008. Currently he is working toward the PhD in ICT from Griffith University (Gold Coast, Australia). Finally, he is interest developing e-learning environments for universities especially in KSA.

**Steve Drew** is a Senior Lecturer in Higher Education at Griffith University and is an adjunct research fellow in the School of Information and Communication Technology. His research interests include acceptance and adoption of information systems, innovative learning environments, and enhanced student engagement with learning. Steve holds a PhD in Computer Science (QUT), and a Masters in Higher Education (Griffith).

**Anne Nguyen** is a Senior Lecturer in the School of Information and Communication Technology at Griffith University and conducts research into a range of e-Systems technologies. Her research interests include e-Commerce, e-Banking, Human-Computer Interactions and User Interface Design.